\documentclass[preprint]{aastex63}

\usepackage{booktabs}
\shorttitle{Active Region Jets} 
\shortauthors{Humphries, Verwichte, Kuridz\'e, Morgan}
\setwatermarkfontsize{240pt}
\usepackage{multimedia}
\graphicspath{{C:/Users/llyrh/Downloads/aastexv6.3/V63/figures/}}

\begin{document}

\title{Multi-wavelength imaging and spectral analysis of jet-like phenomena in a solar active region using IRIS and AIA}
\correspondingauthor{Humphries, Verwichte, Kuridz\'e, Morgan}
\email{llh18@aber.ac.uk, erwin.verwichte@warwick.ac.uk, dak21@aber.ac.uk, hum2@aber.ac.uk}

\author[0000-0002-0786-7307]{Ll\^yr Dafydd Humphries}
\affiliation{Aberystwyth University \\
Faculty of Business and Physical Sciences\\
Aberystwyth, Ceredigion, SY23 3FL, Wales, UK}

\author[0000-0002-0786-7307]{Erwin Verwichte}
\affiliation{Warwick University \\
Centre for Fusion, Space and Astrophysics\\
Coventry, CV4 7AL, England, UK}

\author[0000-0002-0786-7307]{David Kuridz\'e}
\affiliation{Aberystwyth University \\
Faculty of Business and Physical Sciences\\
Aberystwyth, Ceredigion, SY23 3FL, Wales, UK}

\author[0000-0002-0786-7307]{Huw Morgan}
\affiliation{Aberystwyth University \\
Faculty of Business and Physical Sciences\\
Aberystwyth, Ceredigion, SY23 3FL, Wales, UK}

\begin{abstract}

High-resolution observations of dynamic phenomena give insight into properties and processes that govern the low solar atmosphere. We present the analysis of jet-like phenomena emanating from a penumbral foot-point in active region (AR) 12192 using imaging and spectral observations from the Interface Region Imaging Spectrograph (IRIS) and the Atmospheric Imaging Assembly (AIA) on board the Solar Dynamics Observatory. These jets are associated with line-of-sight (LoS) Doppler speeds of $\pm$ 10-22 km s$^{-1}$ and bright fronts which seem to move across the Plane-of-Sky (PoS) at speeds of 23-130 km s$^{-1}$. Such speeds are considerably higher than the expected sound speed in the chromosphere. The jets have signatures which are visible both in the cool and hot channels of IRIS and AIA. 
Each jet lasts on average 15 minutes and occur 5-7 times over a period of 2 hours. Possible mechanisms to explain this phenomenon are suggested, the most likely of which involve p-mode or Alfv\' en wave shock trains impinging on the transition region (TR) and corona as a result of steepening photospheric wavefronts or gravity waves.

\end{abstract}

 
\section{Introduction} \label{sec:intro}

\begin{figure*}[!]
\centering

\includegraphics[trim={2cm 3cm 0 7cm},clip,width=\textwidth]{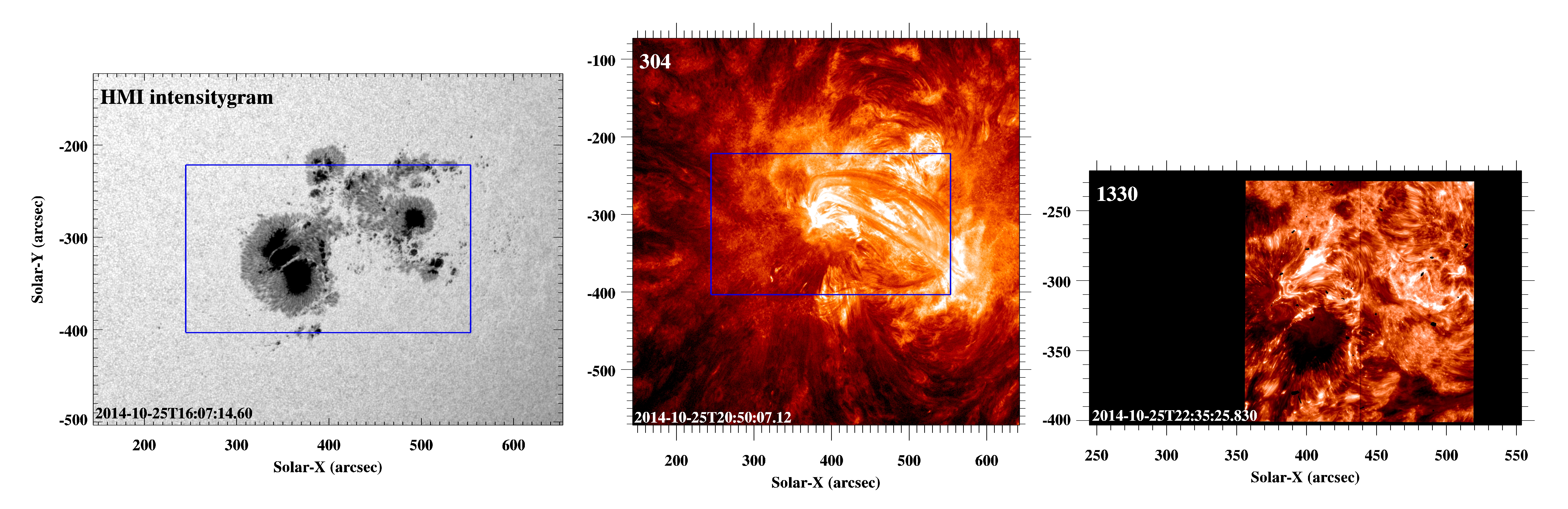}
\caption{HMI intensitygram (left), AIA 304 \AA$\,$ (center) and IRIS 1330 \AA$\,$ (right) (data set B) reference images of the region of interest. The blue rectangle in the HMI and AIA images denotes the IRIS field of view from data set B.}
\label{fig:full_comp}
\end{figure*}

\begin{figure*}[!]
\begin{center}
\includegraphics[trim={4cm 0 0 0},clip,width=\textwidth]{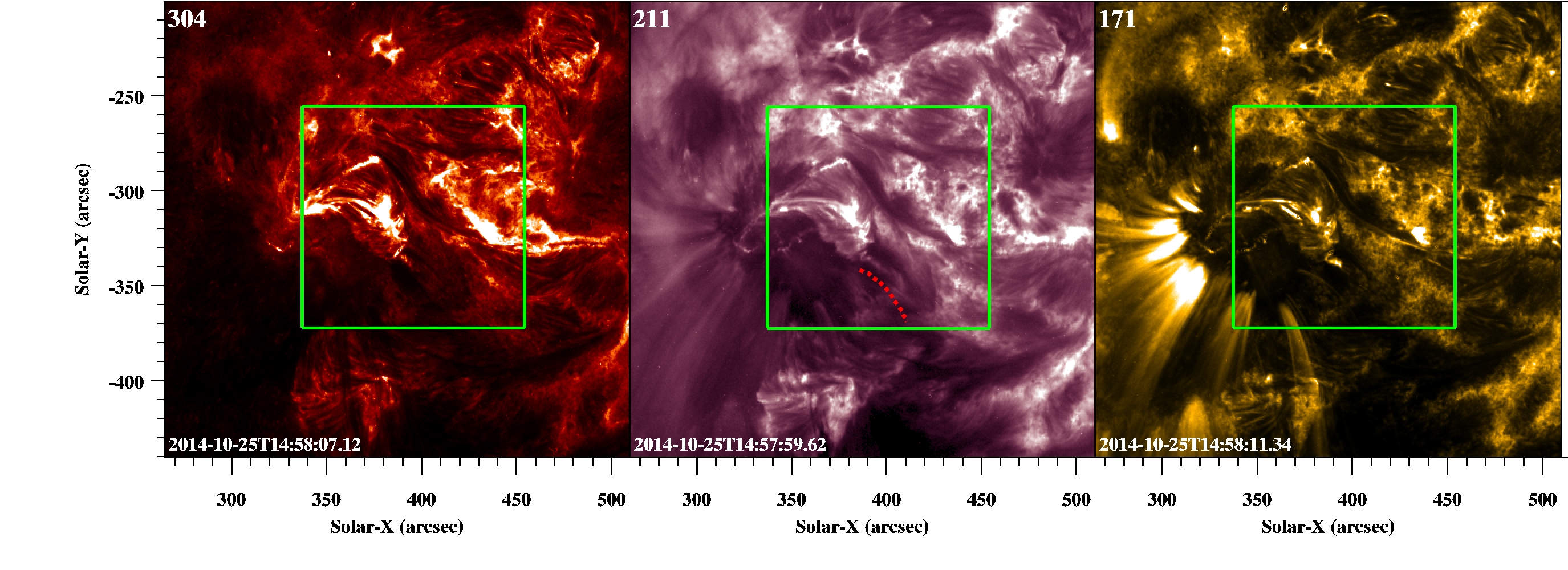}
\caption{AIA 304 \AA, 211 \AA$\,$ and 171 \AA$\,$ images of the region of interest. The green square denotes the IRIS slit-jaw field of view from data set A. The red dotted path denotes an example of a traced jet trajectory used to plot time-distance graphs. 
An animation of this figure is available. The video begins on Oct. 25, 2014 at approximately 14:58:00. The video ends the same day around 17:45:00. The realtime duration is 75 seconds.}
\label{fig:AIA ref afternoon}
\end{center}
\end{figure*}

\begin{figure*}[!]
\begin{center}
\includegraphics[trim={4.5cm 0 0 0},clip,width=\textwidth]{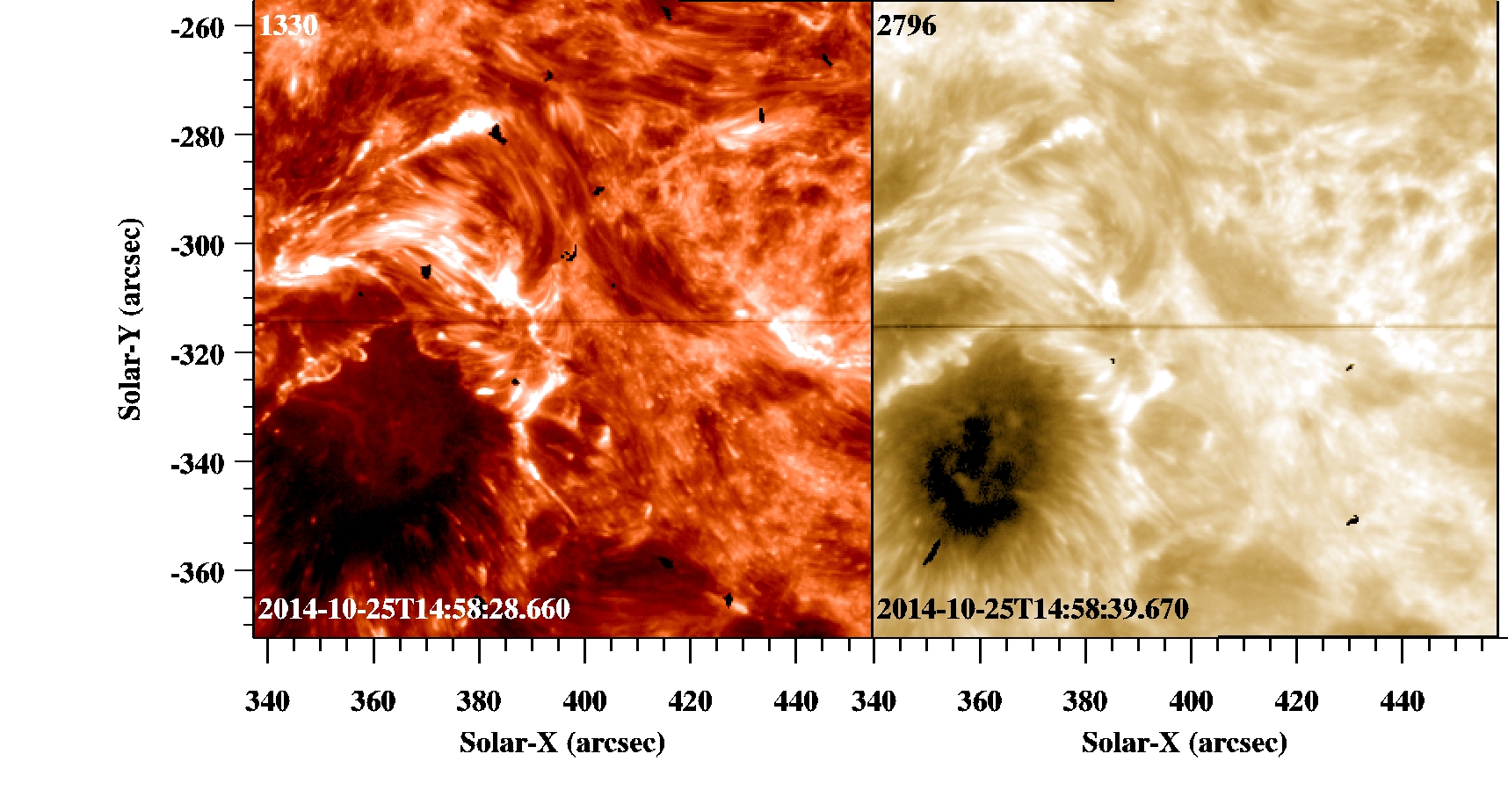}
\caption{IRIS 1330 \AA$\,$ and 2796 \AA$\,$ slit-jaw images from data set A of the region of interest (corresponding to the green box from figure \ref{fig:AIA ref afternoon}). 
An animation of this figure is available. The video begins on Oct. 25, 2014 at approximately 14:58:35. The video ends the same day around 18:00:00. The realtime duration is 34 seconds.
}
\label{fig:IRIS ref afternoon}
\end{center}
\end{figure*}

\begin{figure*}[!]
\begin{center}
\includegraphics[trim={1.5cm 1cm 0 1cm},clip,width=\textwidth]{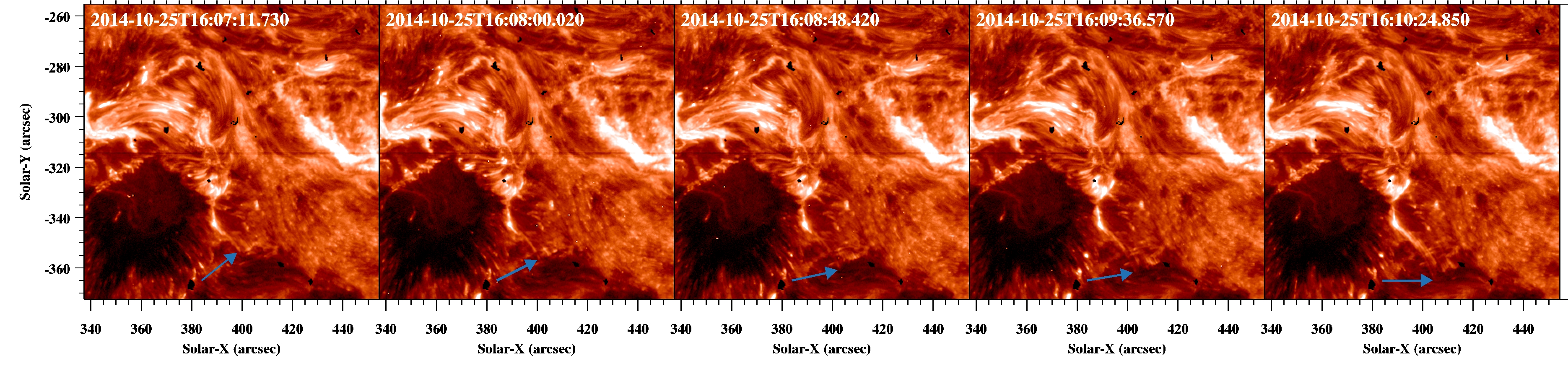}
\caption{IRIS 1330 \AA$\,$ slit-jaw images from data set A showing the progression of a jet front, indicated by the blue arrows. The tip of the arrow indicates the position of the jet front.}
\label{fig:IRIS arrow}
\end{center}
\end{figure*}

The chromosphere is host to a menagerie of dynamic phenomena, including (but not limited to) spicules, jets, shocks and magneto-acoustic waves. These phenomena are interesting as their formation within this physically complicated region, as well as their interaction with other atmospheric layers, particularly the transition region and corona, are not well understood.\\
Fan-shaped jets, ``Peacock'' Jets \citep{robustini} or ``light walls'' \citep{yang} are an example of rarely observed chromospheric phenomenon. These jets are typically observed over sunspot light-bridges \citep{asai} with a mean lifetime of 10 minutes and some intimation in transition region and coronal temperature observations. First observed and analysed by \cite{roy}, fan-shaped jets can extend to 7-50 Mm with driving velocities up to 175 km s$^{-1}$; an order of magnitude greater than the typical $\sim$8 km s$^{-1}$ chromospheric sound speed \citep{anderson}. Their multi-thermal nature is clear, as signatures of these jets have been observed in both IRIS FUV and AIA EUV channels \citep{hou}.\\
Magnetic reconnection is commonly attributed as the fan-shaped jets' driver, typically as a product of the submergence of magnetic fields via convection \citep{lagg} or rapid perpendicular changes between light and dark penumbral fields \citep{jiang}. This seems an appropriate energy release mechanism for emergence speeds in excess of 100 km s$^{-1}$ \citep{louis}. 
Conversely, it has been suggested that type-{\sc{i}}  spicules driven by magneto-acoustic p-mode shocks can manifest as jets \citep{hansteen, zhang}. These spicules typically display upward speeds of 20-25 km s$^{-1}$ \citep{sterling}, which are of comparable magnitude to velocity ranges of jets observed in Ca {\sc{ii}} H lines \citep{shibata_2007}.
\cite{tian} have observed two distinct types of surges above a sunspot light bridge which they determine to be: type-{\sc{i}}  spicules, which are short-lived (approx. 500 s), and; type-{\sc{ii}} spicules, which are impulsive (initial speeds of 50-400 km s$^{-1}$). These phenomena have been shown to reach heights of nearly 10 Mm \citep{beckers}.
\cite{zhang} suggests that these type-{\sc{i}}  spicule observations are the result of p-modes or slow-mode magneto-acoustic waves propagating towards the chromosphere from below. 
Considering these jet's proximity to the active region's sunspot, p-modes could be related to the driving mechanism of running penumbral waves and/or umbral oscillatory events \citep{voort}, which have been observed to exhibit recurrent behaviour on time scales of 3-5 minutes \citep{priya_2017}. 
For example, penumbral waves are produced by photospheric magneto-acoustic activity steepening into shocks at the chromosphere \citep{felipe10} and are primarily observed in Ca {\sc{ii}} line cores with three-minute periodicities \citep{havnes, felipe19}.\\
\cite{hollweg} proposed that spicules or spicule-like structures can be generated by a train of rebound shock fronts repeatedly impacting the transition region from below. These shock trains are a nonlinear development of oscillations of the solar atmosphere at its natural frequency, whereby the initial impulse takes place at the photosphere \citep{rae}.
Plasma falls due to gravity after having been displaced upward by an initial Alfv\' en wavefront. The falling material compresses the atmospheric material below it. This compressed material then rebounds; a process which repeats several times, giving rise to the oscillating wake \citep{stein}. 
These repeating wavefronts steepen into a shock train that is channelled along magnetic field lines. \cite{hollweg_2} elaborated on \cite{hollweg}'s proposition, suggesting that the shock train's repeated impingement results in an approximately constant upward velocity of the transition region and the heating of chromopsheric material. Appreciable transition region and chromospheric motions still occur due to acoustic-gravity waves (driven by Alfv\' en waves) even when considering radiative losses \citep{mariska}.
\\
In section 2, we describe the observational data from AIA and IRIS of recurrent fan-shaped jet phenomena at a sunspot's edge. In section 3, we outline our data analysis methods and the subsequent results. We discuss these results in section 4, and suggest plausible mechanisms to describe the jets' dynamics.

\begin{figure*}[!p]
\includegraphics[trim={3cm 1cm 0 13cm},clip,width=\textwidth]{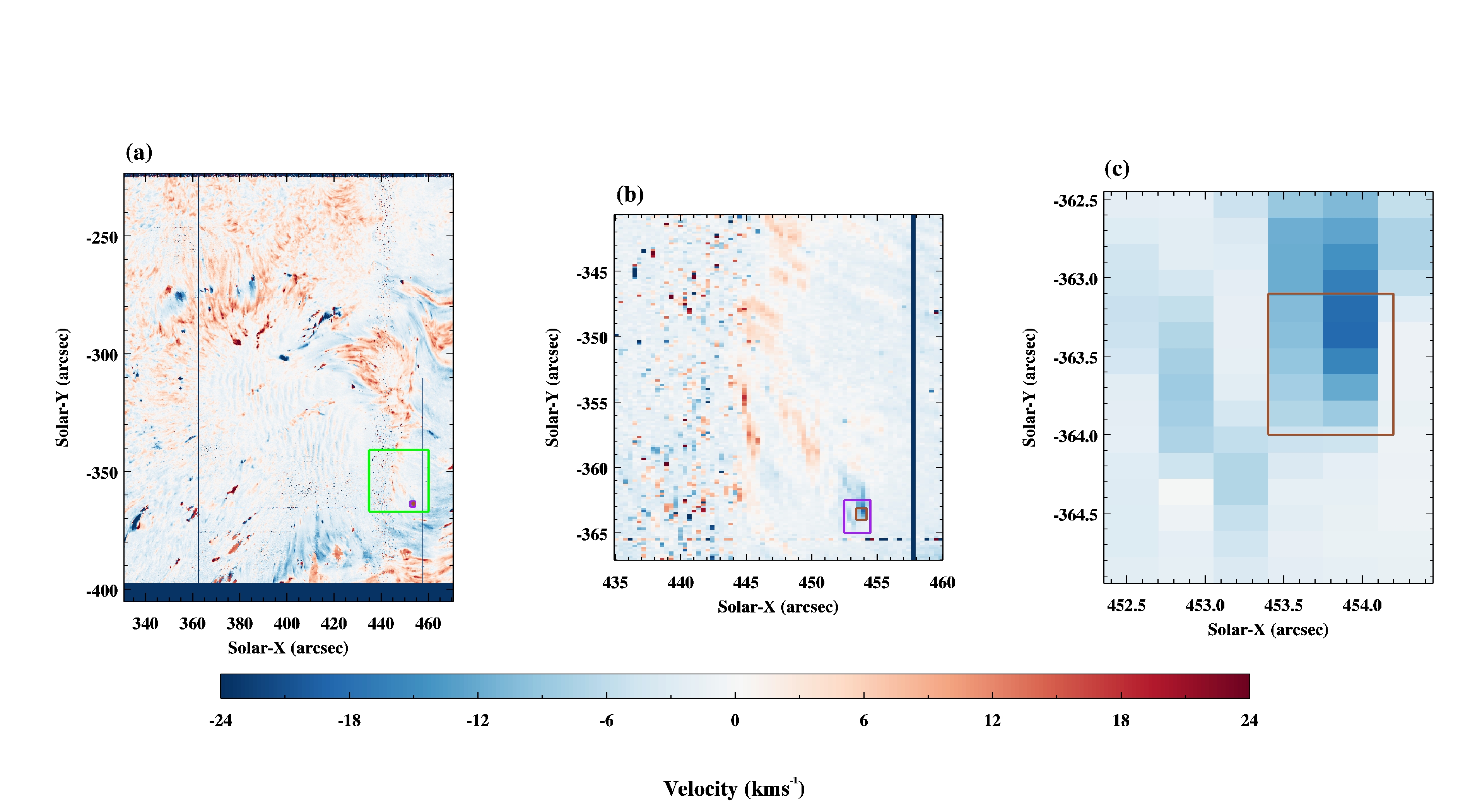} 
\caption{C {\sc{ii}} 1334 Doppler velocity maps gained from data set B showing increasing detail from left to right: (a) - full spectral field of view; (b) - green box from (a); (c) - inset 1 (purple box from (b)). Brown box denotes inset 2; a region of 12 pixels focused on the jet fronts.}
\label{fig:C II vel maps}
\end{figure*}

\begin{figure*}[!p]
\includegraphics[trim={3cm 1cm 0 13cm},clip,width=\textwidth]{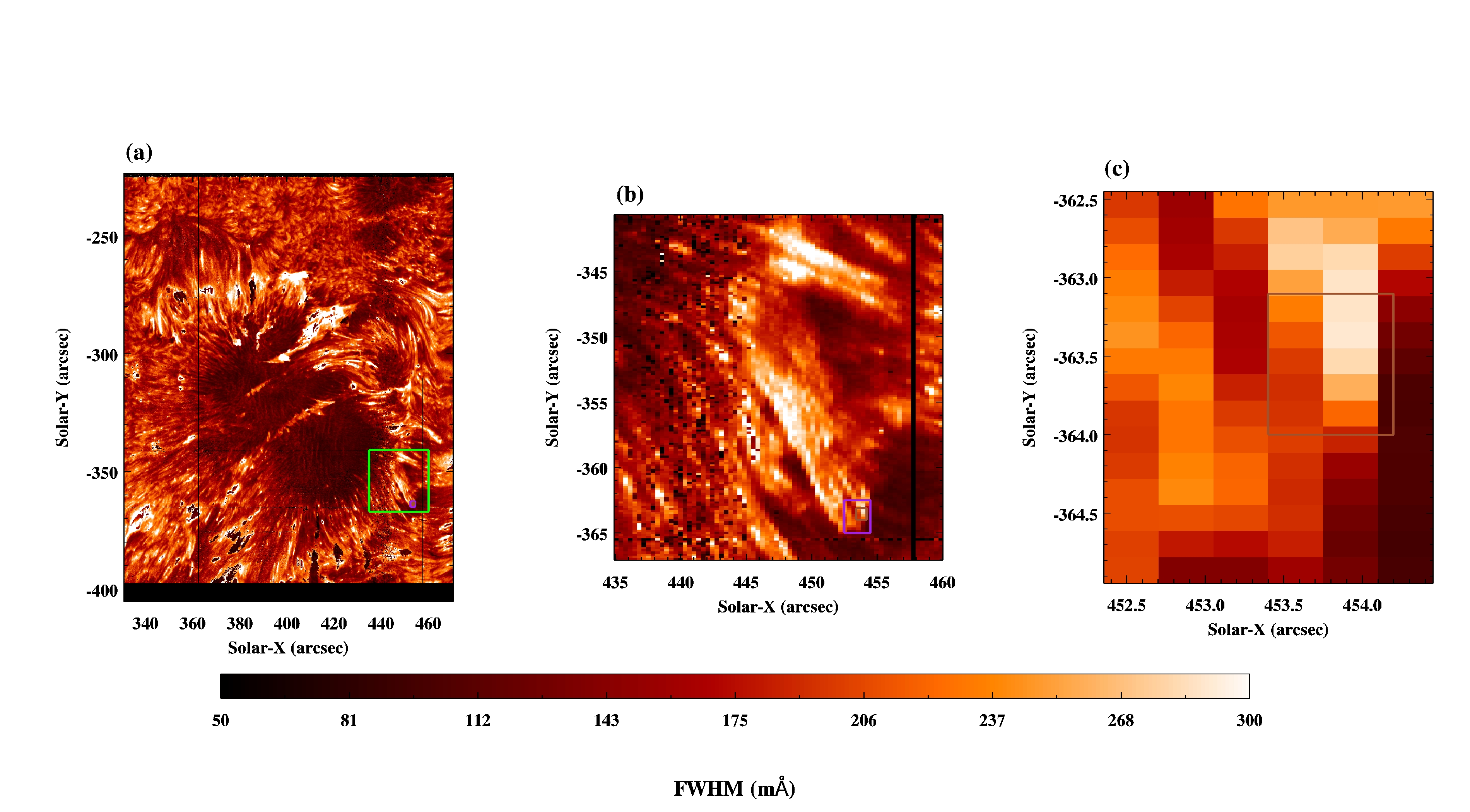}
\caption{C {\sc{ii}} 1336 FWHM maps gained from data set B. See figure \ref{fig:C II vel maps} for details of coloured boxes.}\label{fig:C II wid maps}
\end{figure*}

\begin{figure*}[!t]
\includegraphics[trim={3cm 1cm 0 13cm},clip,width=\textwidth]{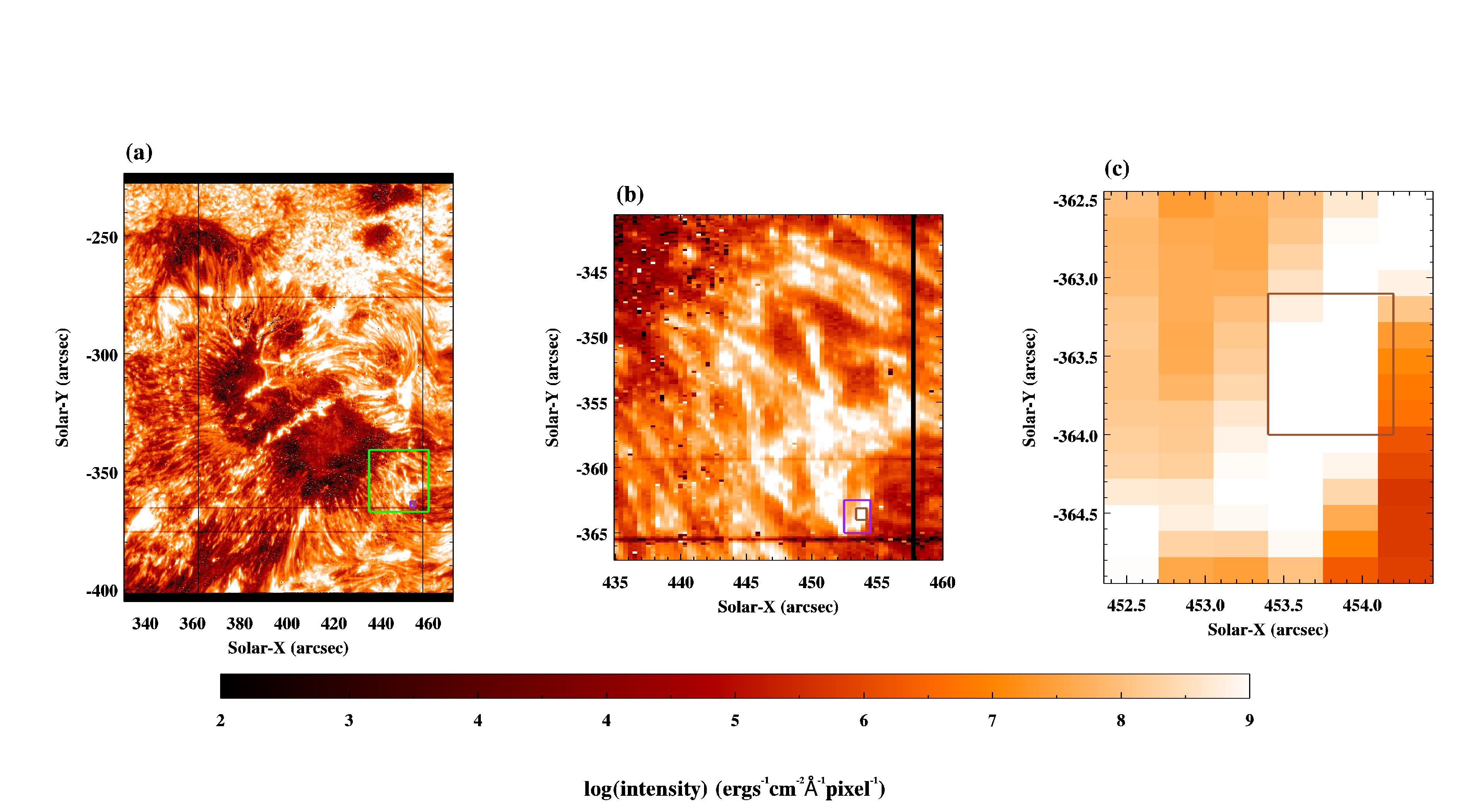}
\caption{Si {\sc{iv}} 1394 Intensity maps gained from data set B. See figure \ref{fig:C II vel maps} for details of coloured boxes.}\label{fig:Si IV int maps}
\end{figure*}

\section{Observations} \label{sec:obs}
We analyse two data sets from AIA \citep{lemen} and IRIS \citep{pontieu} centred on active region 12192. They took place on 25th October 2014 at 14:58:28 UT (data set A) and at 21:00:00 UT (data set B). At that time, the active region is located at the South-West corner of the solar disk (approximately at $X = 400\arcsec, Y = -315\arcsec$). Data set B includes spectroscopic data from IRIS. We have analysed the C {\sc{ii}} and Si {\sc{iv}} lines in detail. Figure \ref{fig:full_comp} shows this active region as a large sunspot complex observed by the Heliosismic and Magnetic Imager \citep[HMI;][]{scherrer}, AIA and IRIS, where the AIA and IRIS images are taken from data set B.\\

The IRIS slit-jaw image data from set A are a large sit-and-stare observation of a flare within the active region, with a 9.4 s temporal cadence, a 0.33$\arcsec$pixel$^{-1}$ spatial resolution and a 121$\arcsec$$\times$128$\arcsec$ field of view. The data consist of two data cubes in the 1330 \AA$\,$ and 2796 \AA$\:$ channels centred on $\sim X=355\arcsec, Y=-303\arcsec$.\\
The IRIS slit-jaw image data from set B are a large, dense raster of the active region's sunspot, with a temporal cadence of 66 s, a spatial resolution of 0.17$\arcsec$pixel$^{-1}$ and a 309$\arcsec$$\times$181$\arcsec$ field of view. The data consist of three data cubes in the 1330 \AA, 1400 \AA, and 2796 \AA$\:$ channels centred on $\sim X=400\arcsec, Y=-315\arcsec$.\\
The IRIS spectroscopic data have a temporal cadence of 6.5 s, a spectral binning of 0.1 \AA, a spatial resolution of 0.17$\arcsec$ and a 120$\arcsec$$\times$119$\arcsec$ field of view (centred on $X=400\arcsec, Y=-315\arcsec$).\\ 
The IRIS data sets were processed using standard procedures to account for the effects of thermal drift, solar rotation and orbital velocity, saturation due to cosmics, and for radiometric calibration.\\
The AIA data from set A have a 12 s temporal cadence and a 0.6$\arcsec$pixel$^{-1}$ spatial resolution. The region of interest covers a 269$\arcsec$$\times$269$\arcsec$ area. The AIA data consists of three data cubes in the 304 \AA, 171 \AA$\,$ and 211 \AA$\,$ channels, each of which are centred at $\sim X=394\arcsec, Y=-322\arcsec$.\\
The region of interest is shown in figure \ref{fig:AIA ref afternoon} which consists of a sunspot with a nearby pore separated by a bright foot-point. Figure \ref{fig:IRIS ref afternoon} shows the IRIS slit-jaw images of the region according to the green box from figure \ref{fig:AIA ref afternoon}. The fronts emerge from this bright foot-point and protrude across the pore. These fronts are clearly visible emanating from the foot-point, as can be seen in figure \ref{fig:IRIS arrow}. Throughout the time series, numerous fronts can be seen emerging from the foot-point before falling back to the source.

\begin{figure*}
\includegraphics[width=\textwidth, height=\textheight]{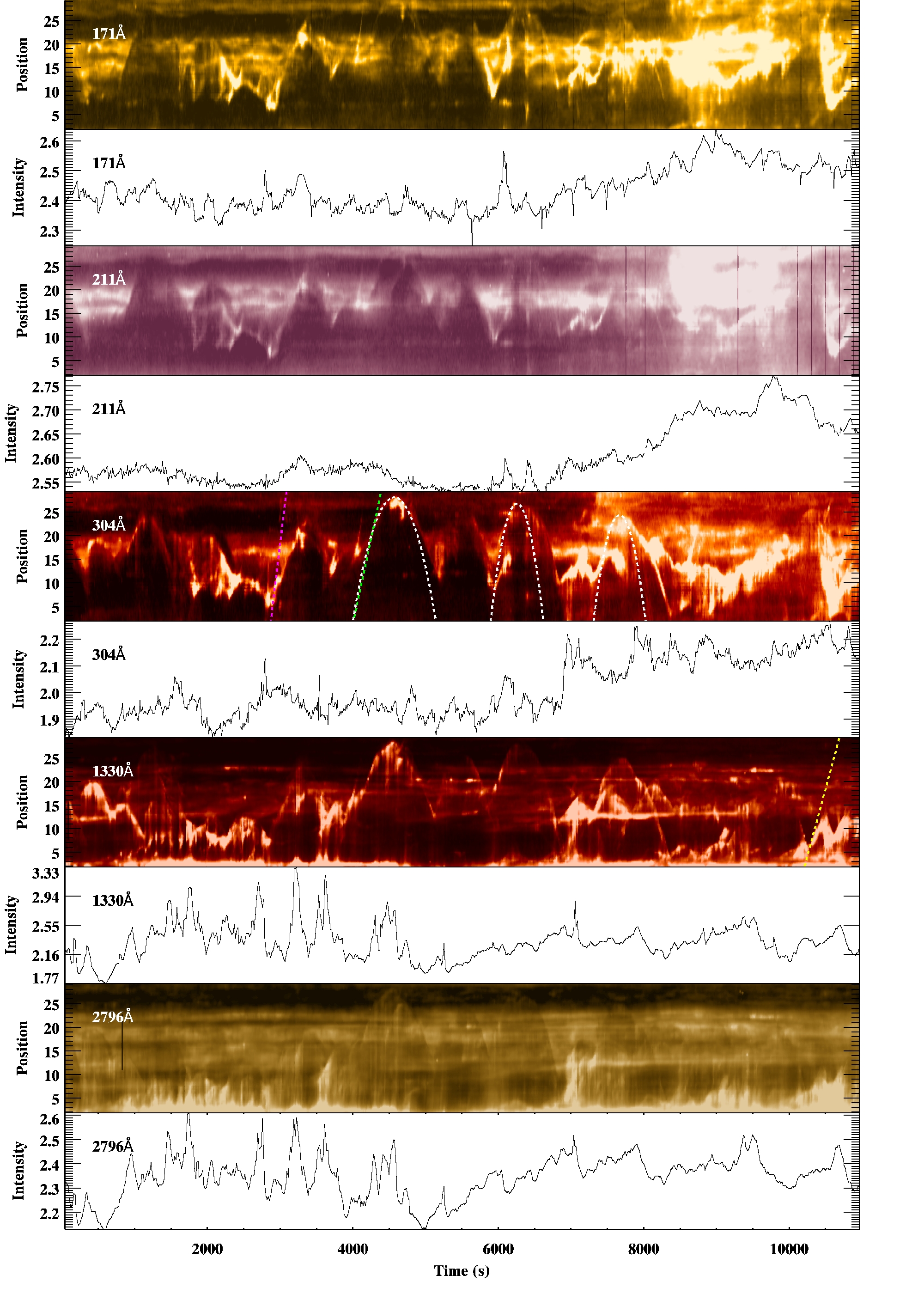}
\caption{AIA and IRIS time-distance graphs of (top-to-bottom) 304, 211, 171, 1330 and 2796$\,$\AA$\,$ set A data according to the red path from figure \ref{fig:AIA ref afternoon}, with their respective over-plotted foot-point light curves. Intensity values are measured in DN. Position values are measured in Mm. White dotted lines indicate parabolic fitting according to jet ascensions. Diagonal dotted lines represent traced gradients of initial velocities: Magenta$=128~km~s^{-1}$; Green$=71.6~km~s^{-1}$; Yellow$=56.5~km~s^{-1}$}\label{fig:curve}
\end{figure*}

\begin{figure*}
\includegraphics[trim={0 0.55cm 0 0},clip,width=\textwidth, height=\textheight]{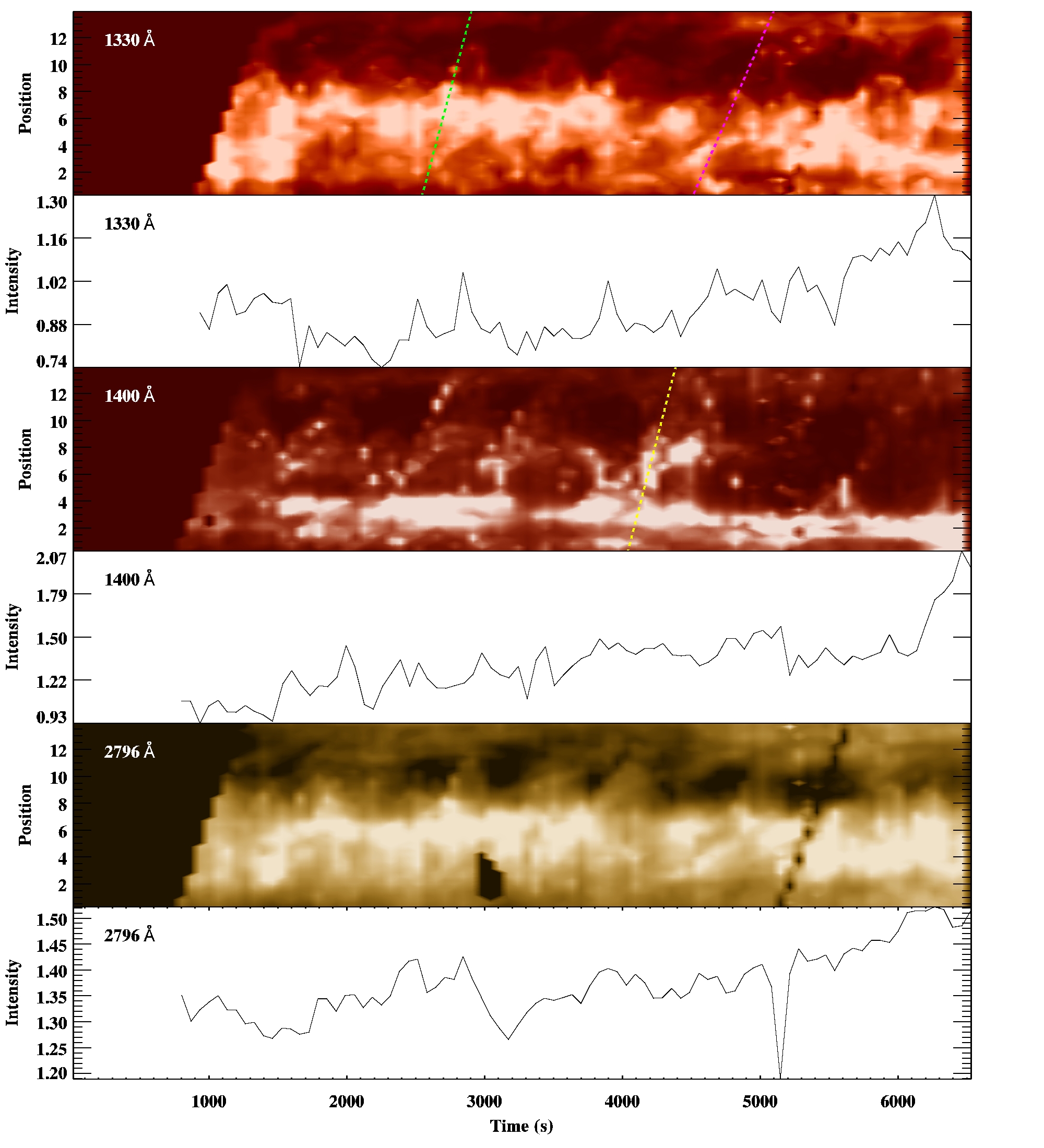} 
\caption{IRIS time-distance graphs of 1330, 1400 and 2796$\,$\AA$\,$ set B data with their respective over-plotted foot-point light curves. These graphs were produced in a similar manner to those from figure \ref{fig:curve}. Intensity values are measured in DN. Position values are measured in Mm. Diagonal dotted lines represent traced gradients of initial velocities as references: Magenta=23.5 km s$^{-1}$; Green=38.5 km s$^{-1}$; Yellow=39.5 km s$^{-1}$}\label{fig:curve_2} 
\end{figure*}

\begin{splitdeluxetable*}{cccccBcccc}
\tabletypesize{\scriptsize}
\tablecaption{Comparison of IRIS spectroscopic ions analysed in this study, their formation temperatures, corresponding atmospheric regions and spectroscopic results\label{tab:1}}
\tablehead{\colhead{ion/wavelength} & \colhead{formation temperature} & \colhead{atmospheric region} & \colhead{LoS Doppler speed (inset 1)} & \colhead{LoS Doppler speed (inset 2)} & \colhead{intensity (inset 1)} & \colhead{intensity (inset 2)} & \colhead{FWHM (inset 1)} & \colhead{FWHM (inset 2)}\\ \colhead{(\AA)} & \colhead{(log\textit{T})} & & \colhead{(km s$^{-1}$)} & \colhead{(km s$^{-1}$)} & \colhead{(erg cm$^{-2}$ s$^{-1}$ sr$^{-1}$)} & \colhead{(erg cm$^{-2}$ s$^{-1}$ sr$^{-1}$)} & \colhead{(\AA)} & \colhead{(\AA)}
}
\startdata
C {\sc{ii}} 1334 & 4.3 & TR & -5.2$\pm 4.2$ & -11.8$\pm 4.3$ & 8000$\pm 5000$ & 16000$\pm 6000$ & 166$\pm 50$ & 215$\pm 35$ \\
C {\sc{ii}} 1336 & 4.3 & TR & -5.6$\pm 4.5$ & -12.7$\pm 4.7$ & 13000$\pm 8000$ & 26000.4$\pm 9500$ & 195$\pm 50$ & 236$\pm 40$ \\
Si {\sc{iv}} 1394 & 4.8 & TR & -6.9$\pm 4.8$ & -15.3$\pm 4.9$ & 8000$\pm 6900$ & 20000$\pm 10000$ & 183$\pm 40$ & 234$\pm 40$ \\
Si {\sc{iv}} 1403 & 4.8 & TR & -8.4$\pm 4.8$ & -16.7$\pm 4.9$ & 3700$\pm 3400$ & 10000$\pm 4800$ & 181$\pm 41$ & 236$\pm 42$ \\
\enddata
\tablecomments{``Inset 1'' denotes the region according to the purple box from figure \ref{fig:C II vel maps}, and ``Inset 2" denotes the region according to the brown box from figure \ref{fig:C II vel maps}. Formation temperatures and solar layers from \cite{pontieu} and \cite{lemen}
}
\end{splitdeluxetable*}

\section{Analysis Methods and Results} \label{sec:method}
\subsection{Spectroscopic Data} \label{sub:spec results}
The IRIS data from set A were a sit-and-stare observation where the spectroscopic slit did not scan over the jets - the position of the slit can be seen in figure \ref{fig:IRIS ref afternoon}. It was therefore necessary to turn to set B's raster IRIS data for spectroscopic analysis. This spectroscopic set B data is used to construct maps of the region of interest according to Doppler/line-of-sight velocities, intensities and FWHM broadenings. These maps are constructed according to the slit position as it moved across the active region during the raster in a similar way to \cite{bamba}.
Figures \ref{fig:C II vel maps}-\ref{fig:Si IV int maps} show velocity, FWHM and intensity maps with progressively smaller fields of view. 
Figure \ref{fig:C II vel maps} shows velocity maps of the region according to the C {\sc{ii}} 1334 line. 
These velocity maps demonstrate Doppler behaviour, whereby red-shift occurs near the foot-point and blue-shift occurs near the top of the front protrusions. 
These red- and blue-shifts are also clear in the C {\sc{ii}} 1336 data (not shown) and less so in the Si {\sc{iv}} 1394 (not shown) and Si {\sc{iv}} 1403 (not shown) data. 
Insets 1 and 2 are chosen as areas for diagnostics due to the presence of this Doppler blue-shifted area in the velocity maps of each spectroscopic ion, the results of which may be seen in Table \ref{tab:1}. 
These values are determined by the mean velocity values within each inset region. The mean Doppler shift range of insets 1 and 2 are -0.9 to -13.2 km s$^{-1}$ and -7.5 to -21.5 km s$^{-1}$, respectively, whereby minus values represent blue-shift. \\
Bifrost simulations show that Doppler shifts of single-Gaussian fits for C {\sc{ii}} 1334 and C {\sc{ii}} 1336 are well correlated with the line-of-sight velocity at unity optical depth, with a Pearson Correlation of 0.88 and 0.83, respectively \citep{rathore}. Therefore these measured shifts provide appropriate values for velocity measurements.\\
Figure \ref{fig:C II wid maps} shows FWHM spectral broadening maps for C {\sc{ii}} 1336$\,$\AA$\,$.
Spectral broadening coincides with the position of bright fronts. They are most prominent in C {\sc{ii}} and less so in Si {\sc{ii}} (not shown). 
Patches of strong broadening activity overlap strong Doppler blue-shift at the insets. 
The mean FWHM spectral broadening in C {\sc{ii}} is 118 \AA$\,$ in inset 1 and 181 \AA$\,$ in inset 2. 
For Si {\sc{iv}} we find 244 \AA$\,$ and 181 \AA, respectively.
Bifrost simulations suggest that FWHM broadenings of the C {\sc{ii}} 1334 data are larger than those determined with a single Gaussian fitting due to the flattened top of the C {\sc{ii}} 1334 profiles, as opposed to broader double-peak profiles. 
Therefore, single-Gaussian fits may be used as an approximate, lower-limit diagnostic of FWHM broadenings for C {\sc{ii}} 1334 \citep{rathore2}. 
It is unclear if the maps (as a stitched-together image of the region of interest) of FWHM broadening follows a single front as it evolves or if FWHM broadening occurs across the entire path as a jet emerges from its foot-point.\\ 
Intensity maps are also constructed, examples of which can be seen in Figure \ref{fig:Si IV int maps} for Si {\sc{iv}} 1394. 
It is clear that there are sharp intensity spikes within the bright front region, particularly within insets 1 and 2, coinciding with the Doppler blue-shifts from the velocity maps. 
Table \ref{tab:1} shows that intensity values for insets 1 and 2 are lowest for Si {\sc{iv}} 1403 (not shown). 
While these intensity data are not analysed in this study, they may be exploited in a future study, providing minimum values of energy output via radiative losses, for example. 

The IRIS raster moves across the region, capturing spectroscopic slices of data at each moment. The velocity, intensity and FWHM maps combine these into spatial images, but there is a time dependence to the sequence. It seems likely that the regions of red-shift and blue-shift are momentary snapshots in which at least one of the fronts are moving away from and towards the observer, respectively. From the data set B slit-jaw images, the slit lies above (or nearly above) the bright foot-point and the source of the fronts at 22:29, or 5800s from the beginning of the observation, and overtakes the fronts at 22:35, or 6300s from the beginning of the observation. This separation of 7.7 minutes between the moment of return and the moment of emergence lies within the range of durations of the most prominent fronts from Figure \ref{fig:curve}.

\begin{deluxetable*}{cccc}
\tablecaption{Comparison of IRIS and AIA imaging channels, their formation temperatures, corresponding atmospheric regions and PoS speeds. The ion and wavelength column denotes each channel analysed as well as its corresponding dominant ion. \label{tab:2}}
\tablehead{\colhead{ion and wavelength} & \colhead{formation temperature} & \colhead{atmospheric region} & \colhead{PoS speed from time-distance results}\\ \colhead{(\AA)} & \colhead{(log\textit{T})} & & \colhead{(km s$^{-1}$)}
}
\startdata
Fe IX 171 & 5.8 & quiet corona, upper TR & $56.5-128.3$\\
Fe {\sc{xiv}} 211 & 6.3 & AR corona & $56.5-128.3$ \\
He {\sc{ii}} 304 & 4.7 & chromosphere, TR & $56.5-128.3$ \\
C {\sc{ii}} 1330 (data set A) & 4.7 & TR & $56.5-128.3$ \\
C {\sc{ii}} 1330 (data set B) & 3.7-7.0 & TR & $23.4-39.5$ \\
Si {\sc{iv}} 1400 & 3.7-5.2 & TR & $23.4-39.5$ \\
Mg {\sc{ii}} h/k 2796 (data set A) & 3.7-4.2 & chromosphere & $56.5-128.3$ \\
Mg {\sc{ii}} h/k 2796 (data set B) & 3.7-4.2 & chromosphere & $23.4-39.5$ \\
\enddata
\end{deluxetable*}

\subsection{Imaging Data} \label{sub:im results}
The jets can be clearly seen as brightenings in the 211, 304, 171, 1330 and 2796 \AA$\,$ data set A images and videos, as well as the 1330, 1400 and 2796 \AA$\,$ data set B images and videos. 
Brightenings are also apparent at the foot-point of the fronts, located at the edge of the sunspot penumbra. 
These foot-point brightenings appear to be recurrent.\\
Figure \ref{fig:curve} shows time-distance graphs constructed along the red dotted path drawn in figure \ref{fig:AIA ref afternoon} for the 304, 211, 171 \AA , 1330 and 2796 channels of data set A. 
The path is traced by using a handful of initial points followed by spline interpolation to create a 100-point path. 
Values found for each of these 100 points are determined using a local average of the surrounding pixels within a 2-pixel radius.
Figure \ref{fig:curve_2} shows time-distance graphs constructed in a similar fashion to those of \ref{fig:curve}, but for the 1330 and 2796 \AA$\,$ IRIS data of set B. 
Data set B took place 6 hours after set A and therefore required a new time-distance path to be plotted. 
A similar path length was used but it was found that set B's jets were far shorter than those observed in set A. 
Therefore it was decided that the distance axis for figure \ref{fig:curve_2} should be reduced accordingly. \\
The jets are evident in data set A's time-distance graphs and are somewhat visible in data set B's time-distance graphs. 
These jets appear as bright fronts which emerge from the foot-point with some initial speed, reach a point of maximum protrusion and then return to the foot-point. 
The absence of initial data of set B's time-distance graphs is due the raster moving over the path; no data are available along the path's coordinates until the raster's field of view begins to pass over the path. 
Additionally, in order to account for IRIS's orbital velocity, Fourier Local Correlation Tracking \citep[FLCT;][]{fw} methods were used to re-align each raster image. 
The nature of the raster, in which it moves over the time-distance path, affects the resolution of the time-distance results. 
This is the reason for the 'blurry' effect in figure 9, rather than any resolution differences between IRIS and AIA images.\\
Only the bright front leading the jet emits in particular wavelengths, while a dark wake trails behind, seen as the dark area beneath the jet curves in data set A's time-distance graphs. 
These dark wakes appear to be as dark as - if not darker than - areas unrelated to the jets. 
Conversely, the area beneath the jets appear bright in set B's time-distance graphs. 
Numerous fronts are present in the graphs with apparent durations of $500-1000s$.\\
Figures \ref{fig:curve} and \ref{fig:curve_2} also include light curves for each of these channels according to the mean intensity from a 2x2 pixel region around the foot-point corresponding to the first point of the path (see figure \ref{fig:curve}). These light curves show recurrent foot-point brightenings that seem to consistently (but not always) occur before the emergence of the rising bright fronts, but can also brighten in the absence of a rising bright front and, in the latter case, seem somewhat unrelated.\\
The diagonal dotted lines are manual tracings of the fronts as they emerge, providing estimates of their emergence speeds. The range of these diagonal lines for both the set A and set B time-distance graphs are outlined in table \ref{tab:2}. 
It appears that the fronts maintain this initial emergence speed for some time before slowing down near the point of maximum protrusion. The fronts then collapse back to their point of origin. The descending portion of the fronts appear to have a less severe gradient, implying a slower return speed. This will be discussed shortly.

\section{Discussion} \label{sec:con}
The results outlined in section \ref{sec:method}, particularly the time-distance graphs, are similar to those of \cite{reid} and \cite{robustini}. The jets develop as bright fronts emanating from a foot-point with temperature signatures ranging from 20000K to near 2MK. 
These fronts may be interpreted as adiabatic plasma heating due to a density increase as the jet strikes upper layers of the solar atmosphere. However, these bright fronts are also present during the jet fall phase. 
The fronts' life-time and frequency are comparable with the jets reported by \cite{reid}. However, we find the fronts protruding a longer distance, i.e. 15-25 Mm compared with 8 Mm reported by \cite{reid}. This could of course be due to projection effects or a greater inclination with respect to the photospheric normal of the magnetic field lines along which the fronts move.
The fronts' evolution also matches previous reports i.e. emerging with some initial speed, continuing with a similar speed until reaching maximum protrusion before returning to, or near to, the foot-point. 
Dark wakes are also evident in \cite{reid}, particularly in the 131$\,$\AA, 171$\,$\AA, 193$\,$\AA$\,$ and faintly in the 335$\,$\AA$\,$ channels. 
These wakes appear darker than the background intensity in coronal bandpasses. 
These dark wakes are likely cool chromospheric material which simply emit far less in these passbands in comparison to the bright fronts, but may still be dense enough to obscure the line-of-sight. \cite{anzer} suggest that these dark wakes are caused by an increased density of Hydrogen and singly-ionised Helium which absorb hot emissions from behind the event along the line-of-sight. 
However, this study also demonstrates 1330$\,$\AA$\,$ and 2796$\,$\AA$\,$ jet signatures, whereby the 2796$\,$\AA$\,$ time-distance data lacks these dark wakes. It could simply be the case that the wakes are an injection of chromospheric material that only emit in the Mg {\sc{ii}} line.
Additionally, the lifetime range of the jets from figures \ref{fig:curve} and \ref{fig:curve_2} (500-1000 s) overlaps with that of type-{\sc{i}}  spicules, which occur on timescales of 180-600s \citep{suematsu}.
\cite{hansteen} and \cite{martinez} have demonstrated, using 1-dimensional and 3-dimensional modelling, respectively, that the top of spicules follow parabolic profiles in time, suggesting that type-{\sc{i}}  spicules typically display quasi-ballistic motion consisting of upward motion followed by downward motion \citep{shibata}, which would seem to coincide with the time-distance results. Set B's jets are evidently shorter and emerge at slower speeds than their set A counterparts.
Finally, the jets' parabolic trajectory appears asymmetric, whereby their ascension gradient is greater than that of their decent.

\subsection{Asymmetries \& peculiarities}
The gradient of the bright fronts' ascension phase appear steeper than their initial gradient of descent, e.g. at times 4500 s, 6300 s, and 8000 s in figure \ref{fig:curve}. 
This is evident from the white dotted parabolic trajectory plots. 
These parabloae were constructed according to plotted points along the jets' ascension phase with the descending half of the parabola plotted symmetrically around the vertex. 
If it the ascending wave were to reflect at the wave front then the gradient of the descent would be more similar to that of the ascent.
In which case, it is possible that the jets' ascent is wave-driven but that the wave pulse will have propagated further up into the atmosphere. 
This would leave the descent to be a mixture of ballistics and the atmosphere's dynamics, whereby the atmosphere contracts to re-establish equilibrium.\\ 
A small decrease in gradient during ascension phases is also visible at the 4500s and 8000s arches. 
While the time-distance graphs in this study represent plane-of-sky trajectories and velocities, it is plausible these drops in speed during ascension are an indicator of ejected material decelerating faster than solar gravitational deceleration alone, provided that the jets' motion consists of more than its line-of-sight component. 
\cite{roy_1973} suggests that a perturbation in the potential field induces a braking force, opposing the jets' emergence, assuming a horizontal magnetic field through which the perturbation tries to move.
However, it is perhaps more likely that the fronts undergo a field-aligned motion and that some bulging of structures may occur due to transverse pressure imbalances. 
This could lead to magnetic pressure and tension forces sufficient to oppose the motion of the jets \citep{verwichte}.\\
Additionally, jets appear brighter in 304, 211, 171 and 1330 during their ascension phase than their descending phase.
\cite{yang} have also observed this phenomenon in light bridges above a sunspot. 
Light wall oscillations do share similar characteristics to this study's jets, such as their duration and oscillation velocity - 3.9 minutes and 15.4 km s$^{-1}$, respectively  - which is of comparable magnitude to the line-of-sight Doppler results from table \ref{tab:1}. However, these light walls appears to be distinct phenomena with maximum lengths of 3.6 Mm as well as \cite{yang}'s IRIS data suggesting that the top and bottom of the wall appear brighter than the body.\\ 

This study’s jets share several similarities with spicules, notwithstanding this study's maximum height results. In particular, their average duration of 5-15 minutes \citep{sterling} and the proposed rapid heating of type-{\sc{ii}} spicules to transition region temperatures \citep{pontieu_2007b} plausibly correlate with this study's time-distance results.
\cite{tian}'s results show that their type-{\sc{ii}} surges can occur in quick succession on timescales of a few minutes, as is the case with the large jets in figure \ref{fig:curve}. 
They also found that the type-{\sc{ii}} phenomena can remain absent for several hours and that significant foot-point brightenings accompany these high-reaching surges. 
This may explain the discrepancy between the speed and length results from set A and set B's time-distance graphs; the maximum foot-point intensity during data set B for each channel is at least 1.7 times fainter than their data set A counterparts.
Specifically, the maximum foot-point brightenings of the set A 1330$\,$\AA$\,$ and 2976$\,$\AA$\,$ data are 2.54 and 1.73 times brighter (in DN) than their set B counterparts, respectively. 
The jets for set B are also shorter in extent and have relatively shorter duration than those of set A. 
Furthermore, the jets of set A show higher initial speeds than those of set B. 
Determining estimates of the set B jets' speeds is difficult due to the spatio-temporal aspect of the scan and the nature of the IRIS raster data; having to co-align these moving frames produces blurry time-distance graphs which inhibit accurate gradient plotting.
\cite{magara} suggests that these type-{\sc{ii}} spicules occur due to magnetic reconnection between a horizontal magnetic flux tube above the light bridge and the background penumbral field. It seems possible that a similar mechanism could occur between the magnetic field lines of a sunspot’s umbra and its penumbra (as is the case in this study’s observations).
Additionally, this may also explain why the density of the area beneath the set B's time-distance arches appears bright compared to those of set A. 
This discrepancy is most likely due to data set B predominantly exhibiting type-{\sc{i}} spicule-like activity occurring more frequently along the drawn path. 
Magnetic reconnection can also account for the multi-thermal nature of this study’s jets: \cite{vissers_2015} suggest that photospheric gas is heated beneath the chromospheric foot-point to momentarily reach temperatures and ionization stages associated with the transition region and corona in a similar manner to \cite{vissers_2013}'s spectroscopic observations of Ellerman bombs.
The Doppler shifts (-11 to -17 km s$^{-1}$) are of a comparable magnitude to those of type-{\sc{i}}  spicules. This is also of a comparable magnitude to the velocity values obtained from data set B's time-distance graphs ($\sim$20-40 km s$^{-1}$).
As previously suggested, the line-of-sight Doppler shifts may not represent real plasma flows; \cite{zaq} observed Doppler shifts in spicules and suggest that these shifts may correspond to the spicule axis moving transversely due to wave propagation within the spicules, as opposed to the actual movement of matter along the spicule axis. 
It is therefore possible that the Doppler shifts represent the amplitude of waves travelling along and within the jet region. A more detailed analysis may help interpret these results.
As a final note on this study's peculiarities, section \ref{sec:obs}'s 2796 $\,$\AA$\,$ video reveals that the nearby umbra demonstrates wave-like brightenings activity. 
These brightenings are typically accompanied by oscillations in the surrounding penumbra. 
These penumbral waves may be linked to the nature of the jet phenomena and a similar methodology to that of \cite{reid} may be used in a subsequent study to determine if such a relationship is present. \\

\subsection{Physical Mechanisms}
\cite{hollweg} demonstrated that shock trains can move the transition region upwards with speeds of 16 km s$^{-1}$, with speeds as large as 50 km s$^{-1}$ at the top of spicules. 
The general tendency for bulk flow speed to increase with height may account for an increase in Doppler shifts with height \citep{beckers} which correlates with \cite{hollweg}'s small or downward velocities near the bottom of the spicule, presumably at the foot-point. This seems to match with the results of this study, whereby blue-shifts of up to 18 km s$^{-1}$ occur near the jets' maximum height while red-shifts tend to occur at the foot-point. 
Whether this study’s Doppler maps show the jets' bulk flow speed, the amplitude of waves within the jets, or momentary snapshots (due to the nature of raster data) of actual line-of-sight motion is currently unclear. 
Regardless, a proposed energy flux of 10$^4$ – 10$^5$ ergs cm$^{-2}$ s$^{-1}$ \citep{hollweg} is paramount if these jets interact with the corona in a similar fashion to spicules. 
Additionally, if these shock front trains are considered to be bi-directional \citep{pasachoff} then a smaller descension gradient (i.e. a slowed return to the foot-point) could be an indicator of gas pressure or some constantly erupting driving force opposing a gravitationally-induced return of material \citep{reid}, which may explain the slowed descent speeds seen in figures \ref{fig:curve} and \ref{fig:curve_2}.\\
Figures \ref{fig:C II wid maps} and \ref{fig:Si IV int maps} demonstrate that both line broadening and intensity increases occur along the entire region of jet activity, following the trajectory of the jets. 
\cite{pontieu2} propose that magneto-acoustic shocks travelling along the line-of-sight can result in non-thermal line broadening and intensity increases of transition region material within regions in which magnetic fields are parallel to the line-of-sight. 
\cite{olluri} also concludes that at least some of the transition region non-thermal line broadening in the presence of shocks with strong velocity gradients (as is the case in data set A's time-distance graphs in \ref{fig:curve}) are caused by slow-mode magneto-acoustic shocks. 
Simulations by \cite{kudoh} suggest that spicule formation (a manifestation of lifting of the transition region), non-thermal broadening of emission lines and even coronal heating are a consequence of Alfv\' en waves (produced within the photosphere) propagating to the corona.
This interpretation may explain the presence of intensity and non-thermal broadening within this study's region of interest with little Doppler shift activity, whereby these Alfv\' en waves effectively churn the plasma as they propagate but demonstrate little discernible line-of-sight plasma motion.\\
More recent work also demonstrates that these rebound shocks can be achieved with a single initial pulse \citep{kuzma, murawski}. \cite{murawski} find that an initial, 30 km s$^{-1}$ velocity-amplitude Alfv\' en pulse – potentially generated due to solar convective motion of magnetic flux tubes \citep{roberts} - leads to typical type-{\sc{i}}  spicule properties such as velocities of 25 km s$^{-1}$ and lengths of 7 Mm. 
Low-chromopsheric Alfv\' en speed estimates assume a density of 10$^{15}$cm$^{-3}$ and a 100G magnetic field strength, yielding a value of 10 km s$^{-1}$ \citep{shibata_1997}. However, while the Alfv\' en velocity requirement is rather high for low-chromospheric speeds, umbral magnetic field strength is at least one magnitude greater than \cite{shibata_1997}'s assumption \citep{norton}.\\
These single-initial-driver models appear to reproduce similar phenomena to this study's jets and are thus worth consideration.
However, models of a periodic initial driver have also re-created spicule properties, such as the leakage and channelling of p-modes along magnetic field lines \citep{pontieu_2007a}. A fall-off of the acoustic cut-off period due to a progressively reduced gravitational influence allows p-modes to tunnel through lower layers and subsequent steepening into magneto-acoustic shocks \citep{pontieu_2004}. This results in an upward motion of chromospheric plasma at $\sim$20 km s$^{-1}$ \citep{pontieu_2005}. \cite{yang} also use this interpretation as the mechanism resulting in their light wall properties. \\
While these magneto-acoustic precursors seem auspicious insofar as they produce similar velocities to this study’s line-of-sight velocities, \cite{robustini} suggests that such a shock-driven initiator is an unlikely mechanism for the plane-of-sky acceleration of these jets, particularly for speeds in excess of 170 km s$^{-1}$ during data set A. However, they may still indirectly result in sufficient jet acceleration as shock waves are often related to quasi-periodic magnetic reconnection \citep{zhang, hillier}.\\

We have investigated recurrent fan-shaped jets protruding from a bright penumbral foot-point. 
These jets occur near-simultaneously with brightenings within their foot-point. The jets appear as bright fronts in AIA's 304 \AA, 211 \AA$\,$ and 171 \AA$\,$ channels as well as IRIS's 1330 \AA$\,$ and 2796 \AA$\,$ channels. 
A darkness is left in the wake of these brightenings in all channels bar 2796 \AA, which is most likely caused by optically thick or cool plasma along the line-of-sight. 
The maximum length of these jets range from $\sim$12-26 Mm with initial plane-of-sky speeds of $\sim$23-130 km s$^{-1}$. The line-of-sight blue-shift velocities of the jet fronts range from $\sim$-1 to -20 km s$^{-1}$. 
FWHM results suggest that broadening occurs over the entire jet region, suggesting that the line-of-sight results may represent the amplitude of waves travelling along the region as opposed to the actual movement of material. 
These results are similar to previous reports of fan-shaped jets but also share similarities with spicule phenomena. Several physical mechanisms are proposed as the initial driver of these phenomena.\\
These channel-specific dark wakes, falling bright fronts and interesting spectroscopic results warrant further investigation into the multi-thermal nature of these jets. Spectral diagnostics of the jet parameters, such as energy, electron density, and temperature using optically thick chromospheric lines, will be considered as avenues for analysis for a subsequent paper.

\acknowledgments

We acknowledge (1) STFC grant ST/S000518/1 to Aberystwyth University which made this work possible, (2) STFC PhD studentship ST/S505225/1 to Aberystwyth University, and (3) a Coleg Cymraeg Cenedlaethol studentship award to Aberystwyth University. 
D.K. has received funding from the S\^{e}r Cymru II scheme, part-funded by the European Regional Development Fund through the Welsh Government and from the Georgian Shota Rustaveli National Science Foundation project FR17 323. 
The SDO data used in this work is courtesy of NASA/SDO and the AIA, EVE, and HMI science teams.
IRIS is a NASA small explorer mission developed and operated by LMSAL with mission operations executed at NASA Ames Research center and major contributions to downlink communications funded by ESA and the Norwegian Space Centre. 
The authors wish to acknowledge FBAPS, Aberystwyth University for the provision of computing facilities and support. 

\vspace{5mm}
\facilities{Aberystwyth University, IRIS, AIA, SDO}

\bibliography{llh18_post_sub}
\bibliographystyle{aasjournal}

\end{document}